\documentclass[12pt, draftclsnofoot, onecolumn]{IEEEtran}

\usepackage{subfigure}
\usepackage{amsmath,amssymb}
\usepackage[dvips]{graphicx}
\usepackage{amsfonts}
\usepackage[mathscr]{eucal}
\usepackage{latexsym}
\usepackage{amsthm}
\usepackage{exscale}
\usepackage[mathscr]{eucal}
\usepackage{bm}
\usepackage[dvipsnames]{color}
\usepackage{cases}
\usepackage{epsfig}
\usepackage[center,small]{caption}
\usepackage{algorithm}
\usepackage{algorithmic}
\usepackage[verbose,nospace,sort]{cite}
\usepackage{tabularx}
\usepackage{multirow}
\usepackage{multicol}
\usepackage{balance}

\graphicspath{{./figsJournal/}}

\scrollmode

\newtheorem{theorem}{Theorem}

\newtheorem{lemma}{Lemma}

\newcommand{\q}{\mathbf q}
\newcommand{\vbd}{\mathbf v}
\newcommand{\abd}{\mathbf a}

\newcommand{\req}{\mathrm{R}}

\newcommand{\ee}{\mathrm{EE}}
\newcommand{\C}{\mathrm{C}}
\newcommand{\rma}{\mathrm{rm}}
\newcommand{\emi}{\mathrm{em}}
\newcommand{\dm}{\mathrm{dm}}
\newcommand{\cir}{\mathrm{cir}}

\newcommand{\lb}{\mathrm{lb}}

\graphicspath{{./figs/}}

\begin{document}
\title{Energy-Efficient UAV Communication with Trajectory Optimization}
\author{Yong~Zeng and Rui~Zhang\\
\thanks{The authors are with the Department of Electrical and Computer Engineering, National University of Singapore (e-mail: \{elezeng, elezhang\}@nus.edu.sg).}
}

\maketitle

\begin{abstract}
Wireless communication with unmanned aerial vehicles (UAVs) is a promising technology for future communication systems. In this paper, we study energy-efficient UAV communication with a ground terminal via optimizing the UAV's trajectory, a new design paradigm that jointly considers both the communication throughput and the UAV's energy consumption. To this end, we first derive a theoretical model on the propulsion energy consumption of fixed-wing UAVs as a function of the UAV's flying speed, direction and acceleration, based on which the energy efficiency of UAV communication is defined. Then, for the case of unconstrained trajectory optimization, we show that both the rate-maximization and energy-minimization designs lead to vanishing energy efficiency and thus are energy-inefficient in general. Next, we introduce a practical circular UAV trajectory, under which the UAV's flight radius and speed are optimized to maximize the energy efficiency for communication. Furthermore, an efficient design is proposed for maximizing the UAV's energy efficiency with general constraints on its trajectory, including its initial/final locations and velocities, as well as maximum speed and acceleration. Numerical results show that the proposed designs achieve significantly higher energy efficiency for UAV communication as compared with other benchmark schemes.
\end{abstract}


\section{Introduction}
Wireless communication by leveraging the use of unmanned aerial vehicles (UAVs) has attracted increasing interest recently \cite{649}. Compared to terrestrial communication systems or those based on high-altitude platforms (HAPs), low-altitude UAV systems are in general more cost-effective by enabling on-demand operations, more swift and flexible for deployment and reconfiguration due to the fully controllable UAV mobility, and are likely to have better communication channels due to the higher chance of line-of-sight (LoS) communication links. 

The main applications of UAV-assisted communications can be loosely classified into three categories \cite{649}. The first one is {\it UAV-aided ubiquitous coverage} \cite{793}, where UAVs are employed to assist the existing terrestrial communication infrastructure, if any, in providing seamless wireless coverage within the serving area. In this case, the UAVs usually stay quasi-stationarily above the serving area acting as aerial base stations (BSs). Two typical scenarios are rapid service recovery after partial or complete infrastructure damage due to natural disasters \cite{615},\cite{794}, and base station offloading in hot spot \cite{795}, which are two important scenarios to be effectively addressed in the fifth-generation (5G) wireless communication systems \cite{613}. Another promising application is {\it UAV-aided relaying} \cite{638,658,641}, where UAVs are despatched to provide reliable wireless connectivity between two or more distant users or user groups in adversary environment, such as between the front line and the command center for emergency responses or military operations. Last but not least, UAV systems could also be employed for {\it UAV-aided information dissemination/data collection} \cite{789}. This is especially appealing for periodic sensing or Internet of Things (IoT) applications, where UAVs could be despatched to fly over the sensors for communications to greatly reduce the sensors' operation power and hence prolong the network lifetime.

Despite their ample applications, UAV communication systems face many new challenges \cite{649}. In particular, the endurance and performance of UAV systems are fundamentally limited by the on-board energy, which is practically finite due to the aircraft's size and weight constraints. Thus, energy-efficient communication for maximizing the information bits per unit energy consumption of the UAV is of paramount importance. Note that energy-efficient designs for UAV communication systems are significantly different from those in the existing literature on terrestrial communication systems \cite{800},\cite{801}. Firstly, while the motivation for energy-efficiency maximization in terrestrial communications is mainly for saving energy consumption and cost, that for UAV systems is more critical due to the limited on-board energy. For example, given the maximum amount of energy that can be carried by the aircraft, an improvement in energy efficiency directly increases the amount of information  bits that can be communicated with the UAV before it needs to be recalled for recharging/refueling. Secondly, besides the conventional energy expenditure on communication-related functions, such as communication circuits and signal transmission, the UAV systems are subject to the additional propulsion power consumption for maintaining the UAV aloft and supporting its mobility (if necessary), which is usually much higher than the communication power consumption (e.g., hundreds of watts versus a few watts). As a result, the communication energy can be even practically ignored compared to the propulsion energy.  Note that the UAV's propulsion energy consumption is determined by its flying status including velocity and acceleration,  which  thus need to be taken into account in energy-efficient design for UAV communications.

In this paper, we study the energy-efficient designs for a point-to-point communication link with a UAV employed to communicate with a ground terminal (GT) for a finite time horizon. The objective is to maximize the energy efficiency in bits/Joule via optimizing the UAV's trajectory, which is a new design framework that needs to jointly consider  the communication throughput and the UAV's propulsion energy consumption. Intuitively, from the throughput maximization perspective, the UAV should stay stationary at the nearest possible location from the GT so as to maintain the best channel condition for communication. However, hovering with strictly zero speed is known to be inefficient (for rotary-wing UAVs) or even impossible (for fixed-wing UAVs) in terms of propulsion energy consumption \cite{766}. Thus, the energy-efficient trajectory design needs to strike an optimal balance between maximizing the communication throughput and minimizing the UAV's propulsion energy consumption. The main contributions of the paper are summarized as follows.
\begin{itemize}
\item First, we derive a theoretical model for the propulsion energy consumption of fixed-wing UAVs as a function of the UAV's flying velocity and acceleration, based on which the energy efficiency of UAV communication is defined. Note that fixed-wing UAVs usually have larger payload and higher speed than their rotary-wing counterparts.\footnote{The analysis in this paper can be extended to rotary-wing UAVs with the energy consumption model modified accordingly.} To the best of our knowledge, this is the first theoretical model that relates the UAV's energy consumption with its velocity (i.e., both flying speed and direction) and acceleration, whereas existing literatures mostly use heuristic energy consumption models only taking into account the speed parameter \cite{792},\cite{791}.
\item Next, for the case of unconstrained UAV trajectory, we study the energy efficiency of the rate-maximization and energy-minimization designs to gain insights. It is shown that the two designs both lead to vanishing energy efficiency, and hence are energy-inefficient in general.
\item We then introduce a practical circular UAV trajectory that is centered at the GT with certain flight radius and speed, which are jointly optimized to maximize the energy efficiency for UAV communication. This result provides a practical design on how a fixed-wing UAV should hover around a GT in order to maximize the communication throughput subject to its limited on-board energy.
\item Furthermore, we study the energy-efficiency maximization problem subject to the general constraints on the UAV's trajectory, including its initial/final locations and velocities, as well as maximum speed and acceleration. An efficient algorithm is proposed to find the approximately optimal trajectory based on linear state-space approximation and sequential convex optimization techniques.
\end{itemize}

Note that energy-efficient UAV communications have been recently studied in e.g., \cite{796}, \cite{798}, but without considering the UAV's propulsion energy consumption. On the other hand, trajectory optimization for UAV communication systems has been studied for various setups. In \cite{658} and \cite{659}, by assuming that the UAV flies with a constant speed, the UAV's heading (or flying direction) is optimized for UAV-assisted wireless relaying and uplink communications, respectively. In \cite{788}, a UAV-based mobile relay is used for forwarding independent data to different user groups. The data volume as well as the relay trajectory in terms of the visiting sequence to the different user groups are optimized based on a genetic algorithm. In \cite{619} and \cite{785}, the deployment/movement of UAVs is optimized to improve the network connectivity of a UAV-assisted ad-hoc network. In our prior work \cite{641}, we study the throughput maximization problem of a UAV-enabled mobile relaying system via joint source/relay power allocation and trajectory optimization. However, none of the above works on trajectory optimization considers the energy efficiency of the system. It is also noted that aircraft trajectory optimization has been studied for other systems not specifically for communication purposes. For instance, mixed-integer linear program (MILP) has been widely applied for trajectory planning for UAV systems to ensure terrain or collision avoidance \cite{620}, \cite{790}. In \cite{791}, the authors study the energy-aware coverage path planning for aerial imaging purposes with the measurement-based energy model of a specific quadrotor UAV. To the authors' best knowledge, this paper is the first work that studies the energy-efficient UAV communication with a generic UAV energy consumption model, which provides a new framework for  designing the UAV trajectory parameters such as its instantaneous velocity and acceleration for communication performance optimization.

The rest of this paper is organized as follows. Section~\ref{sec:SystemModel} introduces the system model and defines the energy efficiency for UAV communication based on a theoretically derived  model on UAV propulsion energy consumption.  In Section~\ref{sec:unConstrained}, the energy efficiency of unconstrained trajectory optimization for rate maximization or energy minimization is studied. Section~\ref{sec:Circular} considers the circular trajectory for energy efficiency maximization. In Section~\ref{sec:EfficientSol}, an efficient algorithm is proposed for the generally constrained trajectory optimization for energy efficiency maximization. Section~\ref{sec:Numerical} presents the numerical results.  Finally, we conclude the paper in Section~\ref{sec:Conclusion}.

{\it Notations: } In this paper, scalars are denoted by italic letters. Boldface lower-case letters denote vectors. $\mathbb{R}^{M\times 1 }$ denotes the space of $M$-dimensional real-valued vector. For a vector $\mathbf a$, $\|\mathbf a\|$ represents its Euclidean norm, and $\mathbf a^T$ denotes its transpose. $\ln(\cdot)$ and $\log_2(\cdot)$ denote the natural logarithm and logarithm with base $2$, respectively. $\tan^{-1}(\cdot)$ is the inverse tangent function. For a time-dependent function $\mathbf x(t)$, $\dot{\mathbf x}(t)$ and $\ddot{\mathbf x}(t)$ denote the derivative and double derivative with respect to time $t$, respectively.

\section{System Model and UAV Energy Efficiency}\label{sec:SystemModel}
\subsection{System Model}
\begin{figure}
\centering
\includegraphics[scale=0.7]{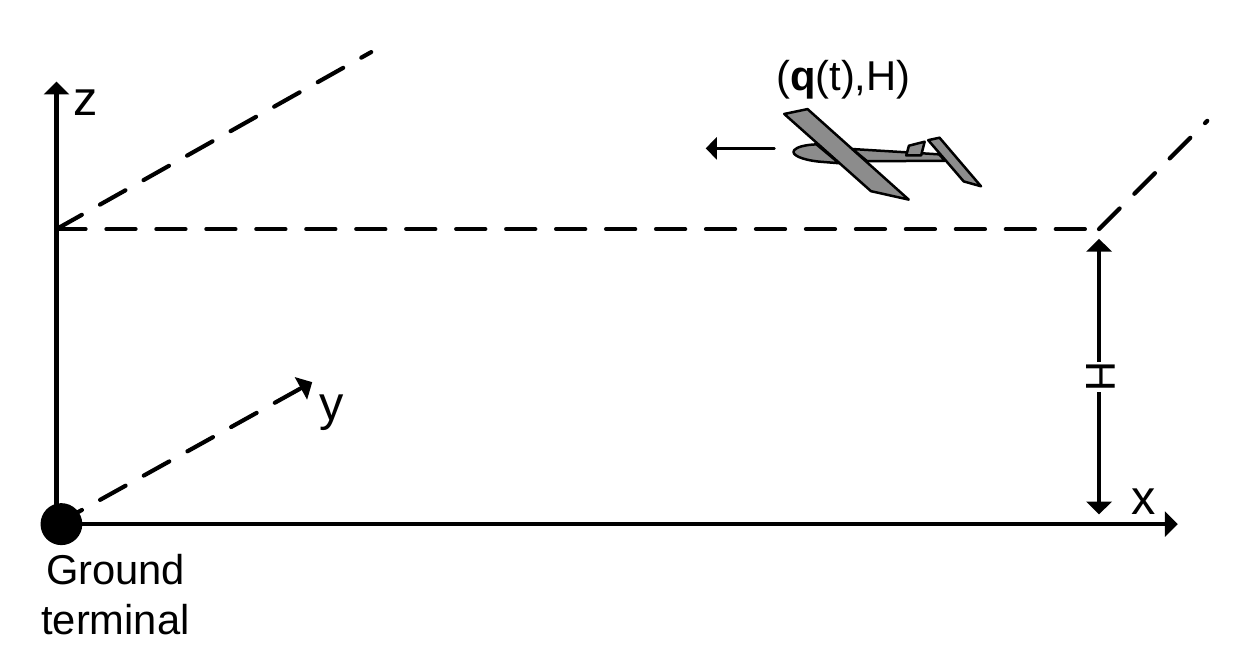}
\caption{Point-to-point wireless communication from a UAV to a ground terminal.}\label{F:SystemSetup}
\end{figure}

As shown in Fig.~\ref{F:SystemSetup}, we consider a wireless communication system where a UAV is employed to send information to a GT. Our objective is to optimize the UAV's trajectory so as to maximize the energy efficiency in bits/Joule for a finite time horizon $T$, where energy efficiency is defined as the aggregated information bits that are transmitted to the GT normalized by the UAV's total energy consumption over the duration $T$.

Without loss of generality, we consider a three-dimensional (3D) Cartesian coordinate system such that the GT is located at the origin $(0,0,0)$. Furthermore, we assume that the UAV flies horizontally at a constant altitude $H$. 
Denote the UAV trajectory projected on the horizontal plane as $\mathbf q(t)=[x(t), y(t)]^T\in \mathbb{R}^{2\times 1}$, where $0\leq t \leq T$. Thus, the time-varying distance from the UAV to the GT can be expressed as
 \begin{align}
 d(t)=\sqrt{H^2+\|\mathbf q(t)\|^2}, \ 0\leq t\leq T.
  \end{align}

  For ease of exposition, we assume that the communication link from the UAV to the GT is dominated by the LoS channel. Furthermore, the Doppler effect due to the UAV mobility is assumed to be perfectly compensated. Therefore, the time-varying channel  follows the free-space path loss model, which can be expressed as
   \begin{align}
   h(t)=\beta_0d^{-2}(t)=\frac{\beta_0}{H^2+\|\q(t)\|^2}, \ 0 \leq t \leq T,
   \end{align}
   where $\beta_0$ denotes the channel power at the reference distance $d_0=1$m. Assuming constant power transmission $P$ by the UAV, the instantaneous channel capacity in bits/second can be expressed as
   \begin{align}
   R(t)= &B \log_2\left(1+\frac{Ph(t)}{\sigma^2}\right)\notag \\
  =& B \log_2\left(1+\frac{\gamma_0}{H^2+\|\q(t)\|^2}\right), \ 0\leq t \leq T,
   \end{align}
   where $B$ denotes the channel bandwidth, $\sigma^2$ is the white Gaussian noise power at the GT receiver, $\gamma_0=\beta_0P/\sigma^2$ is the reference received signal-to-noise ratio (SNR) at $d_0=1$m. Thus, the total amount of information bits $\bar R$ that can be transmitted from the UAV to the GT  over the duration $T$ is a function of the UAV trajectory $\q(t)$, expressed as
   \begin{align}\label{eq:Rbar}
   \bar R\big(\q(t)\big)=\int_0^T B\log_2\left(1+\frac{\gamma_0}{H^2+\|\q(t)\|^2}\right) dt.
   \end{align}

\subsection{UAV Energy Consumption Model and Energy Efficiency}
   The total energy consumption of the UAV includes two components. The first one is the communication-related energy, which is due to the radiation, signal processing, as well as other circuitry. The other component is the propulsion energy, which is required for ensuring that the UAV remains aloft as well as for supporting its mobility, if needed. Note that in practice, the communication-related energy is usually much smaller than the UAV's propulsion energy, and thus is ignored in this paper. Furthermore, as shown in Appendix~\ref{A:energyModel}, for fixed-wing UAVs under normal operations, i.e., no abrupt deceleration that requires the engine to abnormally produce a reverse thrust against the forward motion of the aircraft, the total propulsion energy required   is a function of the trajectory $\q(t)$, which  is expressed as
\begin{align}
\hspace{-1ex}\bar E(\q(t))= & \int_{0}^T \hspace{-1ex} \left[c_1\|\mathbf v(t)\|^3+\frac{c_2}{\|\mathbf v(t)\|}\bigg(1+\frac{\|\mathbf a(t)\|^2-\frac{(\mathbf a^T(t) \mathbf v(t))^2}{\|\mathbf v(t)\|^2}}{g^2}\bigg)\right]dt\notag \\
&+\frac{1}{2}m \left(\|\mathbf v(T)\|^2-\|\mathbf v(0)\|^2 \right),\label{eq:ER}
\end{align}
where
\begin{align}
\mathbf v(t)\triangleq \dot{\q}(t), \ \mathbf a(t)\triangleq \ddot{\q}(t)
 \end{align}
 denote the instantaneous UAV velocity and acceleration vectors, respectively, $c_1$ and $c_2$ are two parameters related to the aircraft's weight, wing area, air density, etc., as expressed in \eqref{eq:c1c2} of Appendix~\ref{A:energyModel}, $g$ is the gravitational acceleration with nominal value $9.8$ m/s$^2$, $m$ is the mass of the UAV including all its payload.\footnote{For simplicity, we ignore the UAV energy storage weight reduction over time as more fuel or battery is consumed.}

 The expression in \eqref{eq:ER} shows that for level flight with fixed altitude,  the UAV's energy consumption only depends on the velocity $\mathbf v(t)$ and acceleration $\mathbf a(t)$, rather than its actual location $\q(t)$. Furthermore,  the result in \eqref{eq:ER} can be interpreted based on the well known work-energy principle. The integral term in \eqref{eq:ER}, which is guaranteed to be positive, is the work required from the aircraft's engine to overcome the air resistance force (or the drag). It depends on the UAV speed $\|\mathbf v(t)\|$, as well as its centrifugal acceleration $a_{\perp}(t)\triangleq \sqrt{\|\mathbf a(t)\|^2-\frac{(\mathbf a^T(t) \mathbf v(t))^2}{\|\mathbf v(t)\|^2}}$, i.e., the acceleration component that is normal to the UAV velocity vector and causing heading (direction) changes, yet without altering the UAV speed. The second term in \eqref{eq:ER}, denoted as $\Delta_K$, represents the change in the UAV's kinetic energy, an aggregated effect of the UAV's tangential acceleration component that is in parallel to the UAV's velocity vector. Thus, $\Delta_K$ only depends on the initial and final speeds $\|\mathbf v(T)\|$ and $\|\mathbf v(0)\|$, rather than the intermediate UAV state. Note that if $\|\mathbf v(t')\|=0$ for some $t'$, then the required energy  $\bar E\rightarrow \infty$, which reflects the fact that a fixed-wing aircraft must maintain a forward motion to remain aloft.

\begin{figure}
\centering
\includegraphics[scale=1.5]{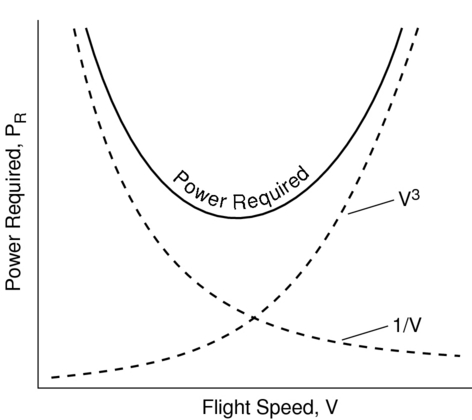}
\caption{Typical power required curve versus speed $V$ for a UAV in straight-and-level flight.}\label{F:PowerSteadyLevel}
\end{figure}

For steady straight-and-level flight (SLF) with constant speed $V$, we have $\|\mathbf v(t)\|=V$ and $\mathbf a(t)=\mathbf 0$, $\forall t$. Thus, \eqref{eq:ER} reduces to
\begin{align}
\bar E_{\mathrm{SLF}}(V)=T \left(c_1 V^3+\frac{c_2}{V}\right).\label{eq:Estd}
\end{align}
The power consumption of \eqref{eq:Estd} as a function of $V$ is illustrated in Fig.~\ref{F:PowerSteadyLevel}, which consists of two terms.  The first term, which is proportional to the cubic of the speed $V$, is known as the {\it parasitic power} for overcoming the  parasitic drag due to the aircraft's skin friction, form drag, etc. The second term, which is inversely proportional to $V$, is known as the {\it induced power} for overcoming the lift-induced drag, i.e., the resulting drag force due to wings redirecting air to generate the lift for compensating the aircraft's weight \cite{766}.

 With \eqref{eq:Rbar} and \eqref{eq:ER}, the energy efficiency of the UAV communication system can thus be expressed as
 \begin{align}
 \ee(\q(t))=\frac{\bar R(\q(t))}{\bar E(\q(t))}.\label{eq:EE}
 \end{align}


   \section{Energy Efficiency with Unconstrained Trajectory}\label{sec:unConstrained}
In this section, we study the energy efficiency of unconstrained UAV trajectory for rate maximization and energy minimization, respectively, to gain insights and provide benchmarks.

\subsection{Rate-Maximization Trajectory}\label{sec:UncontrRM}
In the absence of any constraint on $\q(t)$, it immediately follows from \eqref{eq:Rbar} that the {\it rate-maximization} (rm) UAV trajectory  should be $\q(t)=\mathbf 0$, $\forall t$, i.e., the UAV should stay stationary just above the GT to maintain the best communication channel. The resulting aggregated information throughput is
\begin{align}
\bar R_{\rma}=TB\log_2\left(1+\frac{\gamma_0}{H^2}\right).
\end{align}
 In this case, since $\mathbf v(t)=\dot{\q}(t)=\mathbf 0$, $\forall t$, the corresponding energy consumption in \eqref{eq:ER} is $\bar E_{\rma}\rightarrow \infty$. Thus, the resulting energy efficiency with the rate-maximization design is
\begin{align}
\ee_{\rma}=\frac{\bar R_{\rma}}{\bar E_{\rma}}=0, \ \forall T>0
\end{align}
which is evidently energy-inefficient.

\subsection{Energy-Minimization Trajectory}\label{sec:UnconstrEM}
Next, we consider the {\it energy-minimization} design, i.e., the UAV trajectory $\q(t)$ is optimized merely for minimizing the total energy consumption without considering the communication performance. By ignoring the change in kinetic energy in the last term of \eqref{eq:ER}, which is independent of the time duration $T$ and hence is practically negligible for large $T$, the energy-minimization trajectory design problem can be formulated as
\begin{align}
\underset{\mathbf v(t), \mathbf a(t)}{\min}\ & \int_{0}^T \hspace{-1ex} \left[c_1\|\mathbf v(t)\|^3+\frac{c_2}{\|\mathbf v(t)\|}\bigg(1+\frac{\|\mathbf a(t)\|^2-\frac{(\mathbf a^T(t) \mathbf v(t))^2}{\|\mathbf v(t)\|^2}}{g^2}\bigg)\right]dt\notag \\
\text{s.t.}\ & \mathbf a(t)=\dot{\mathbf v}(t), \ \forall \  0\leq t \leq T. \label{eq:powerMimum}
\end{align}

\begin{theorem}\label{theo:energyMini}
The optimal solution to the energy-minimization (em) problem \eqref{eq:powerMimum} is
\begin{align}
\mathbf a^\star(t)=\mathbf 0, \ \mathbf v^\star(t)=V_{\emi}\vec{\mathbf v},\ \forall t,\label{eq:optEM}
\end{align}
where $V_{\emi}\triangleq \left(c_2/(3c_1)\right)^{1/4}$ is the energy-minimum speed and $\vec{\mathbf v}$ is an arbitrary unit-norm vector denoting the constant UAV flying direction. The corresponding minimum energy consumption is
\begin{align}
\bar E_{\emi}=P_{\min} T,\label{eq:Emi}
\end{align}
with $P_{\min}\triangleq \left(3^{-3/4}+3^{1/4}\right)c_1^{1/4}c_2^{3/4}$ denoting the minimum power consumption of the UAV.
\end{theorem}

\begin{IEEEproof}
Please refer to Appendix~\ref{A:energyMini}.
\end{IEEEproof}

Theorem~\ref{theo:energyMini} shows that for the unconstrained trajectory optimization, the energy-minimization solution is simply the steady straight-and-level flight  with the power-minimum speed $V_{\emi}$. Note that the energy-minimization trajectory is non-unique, since the initial UAV location $\q(0)$ and the flying direction $\vec{\mathbf v}$ can be arbitrary. This thus gives us the degree of freedom to find the best energy-minimization trajectory that gives the maximum aggregated information throughput. It is not difficult to show that among all the energy-minimization trajectories, i.e., all straight-and-level flights with the energy-minimum speed $V_{\emi}$, those being symmetric around the GT (or the origin) lead to the highest information throughput. Therefore, without loss of optimality, the trajectory can be expressed as $\q(t)=[x(t), 0]^T$, where
\begin{align}
x(t)=-\frac{V_{\emi}T}{2}+V_{\emi}t, \ 0\leq t \leq T.
 \end{align}
 By substituting $\q(t)$ into \eqref{eq:Rbar}, the maximum aggregated information throughput by the energy-minimization trajectory design can be expressed as
\begin{align}
&\bar R_{\emi}= B\int_0^T \log_2\left(1+\frac{\gamma_0}{H^2+(-\frac{V_{\emi}T}{2}+V_{\emi}t)^2}\right)dt \notag \\
=&\frac{4B}{(\ln 2)V_{\emi}}\bigg[\frac{V_{\emi}T}{4} \ln \left( 1+\frac{\gamma_0}{H^2+(\frac{V_{\emi}T}{2})^2}\right)
+\notag \\
&\sqrt{H^2+\gamma_0}\tan^{-1}\left(\frac{V_{\emi}T}{2\sqrt{H^2+\gamma_0}}\right)
-H\tan^{-1}\left(\frac{V_{\emi}T}{2H}\right) \bigg],\label{eq:Remi}
\end{align}
where \eqref{eq:Remi} follows from the change of variable $z=-\frac{V_{\emi}T}{2}+V_{\emi}t$ and we have used  the integral formula given in \eqref{eq:integralFormula} shown on the top of the next page.
\begin{figure*}
\begin{equation}
\small
\begin{aligned}\label{eq:integralFormula}
F(z)\triangleq \int &\ln\left(1+\frac{\gamma_0}{H^2+z^2}\right)dz=
z \ln \left( 1+\frac{\gamma_0}{H^2+z^2}\right)
+2\sqrt{H^2+\gamma_0}\tan^{-1}\left(\frac{z}{\sqrt{H^2+\gamma_0}}\right)
-2H\tan^{-1}\left(\frac{z}{H}\right)
\end{aligned}
\end{equation}
\hrule
\end{figure*}





As a result, the maximum energy efficiency achievable by the energy-minimization trajectory design can be obtained in closed-form as
$\ee_{\emi}=\bar{R}_{\emi}/\bar E_{\emi}$.
For sufficiently large operation duration $T\rightarrow \infty$, the aggregated information throughput in \eqref{eq:Remi} reduces to
\begin{align}
R_{\emi}\rightarrow \frac{2\pi B}{(\ln 2)V_{\emi}} \left(\sqrt{H^2+\gamma_0}-H\right),  \text{ as } T\rightarrow \infty,
\end{align}
which is finite and independent of $T$.
On the other hand, as the minimum energy consumption in \eqref{eq:Emi} linearly increases with $T$, the resulting energy efficiency is thus
\begin{align}
\ee_{\emi} \rightarrow 0, \text{ as } T\rightarrow \infty.
\end{align}
Thus, the energy-minimization trajectory design is also energy-inefficient in general.

\begin{figure}
\centering
\includegraphics[scale=0.8]{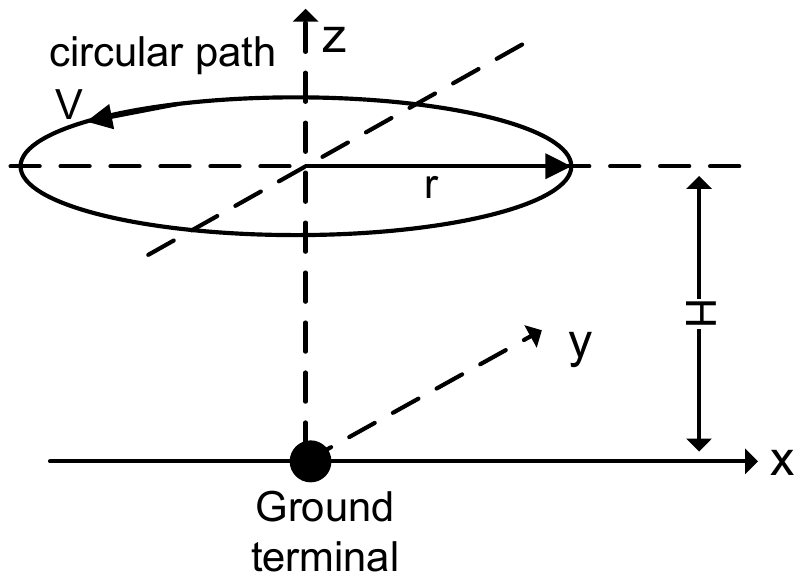}
\caption{An illustration of the steady circular flight.}\label{F:CircularPath}
\end{figure}

\section{Energy-Efficiency Maximization with Circular Trajectory}\label{sec:Circular}
The preceding section shows that neither the rate-maximization nor the energy-minimization trajectory design is  energy-efficient. In general, energy-efficient trajectory design needs to achieve an optimal tradeoff between these two objectives. 
In this section,  we propose a practical energy-efficient  design by assuming that the UAV follows a circular trajectory centered at the GT with constant speed $V$ and radius $r$, as illustrated in Fig.~\ref{F:CircularPath}. Intuitively, a smaller radius $r$, though achieving higher information throughput with the GT, also consumes more power by the UAV to maintain a more acute heading change, and vice versa. Therefore, both $V$ and $r$ need to be optimized for maximizing the energy efficiency.

With circular (cir) trajectory, it follows from \eqref{eq:Rbar} that the aggregated information throughput reduces to a function of the radius $r$ as
\begin{align}
\bar R_{\cir}(r)=TB\log_2\left(1+\frac{\gamma_0}{H^2+r^2}\right).\label{eq:Rcir}
\end{align}
Note that $\bar R_{\cir}(r)$ is maximized when $r=0$, i.e., the rate-maximization circular flight reduces to the extreme case of stationary hovering as studied in Section~\ref{sec:UncontrRM}, which is known to be energy-inefficient.

On the other hand, for steady circular flight with constant speed $V$, we have
$\|\mathbf v(t)\|=V$ and $\mathbf a^T(t)\mathbf v(t)=0$, $\forall t$, i.e., the acceleration (if any) must be perpendicular to the velocity to ensure no speed variation.  Furthermore, the centrifugal acceleration for maintaining the circular path is known to be proportional to the speed square $V^2$ and inversely proportional to the circle radius $r$, i.e., $\|\mathbf a(t)\|=V^2/r, \ \forall t$. 
As a result, the UAV energy consumption \eqref{eq:ER} with steady circular flight reduces to
\begin{align}
\bar E_{\cir}(V,r)=T\left[\Big(c_1+\frac{c_2}{g^2r^2}\Big)V^3+\frac{c_2}{V}\right].\label{eq:Ecir}
\end{align}
It is not difficult to find that $\bar E_{\cir}(V,r)$ is minimized when $r\rightarrow \infty$ and $V=V_{\emi}$, i.e., the energy-minimization circular flight reduces to the extreme case of straight flight with energy-minimum speed $V_{\emi}$, which has also been shown to be energy-inefficient in Section~\ref{sec:UnconstrEM}.

With \eqref{eq:Rcir} and \eqref{eq:Ecir}, the energy efficiency in \eqref{eq:EE} for circular trajectory reduces to
\begin{align}\label{eq:eeCir}
\ee_{\cir}(V,r)=\frac{\bar R_{\cir}(r)}{\bar E_{\cir}(V,r)}=\frac{B\log_2\left(1+\frac{\gamma_0}{H^2+r^2}\right)}{\Big(c_1+\frac{c_2}{g^2r^2}\Big)V^3+\frac{c_2}{V}}.
\end{align}
Note that the energy efficiency in \eqref{eq:eeCir} is independent of $T$. The energy-efficiency maximization problem can then be formulated as
\begin{align}
\underset{V\geq 0, r\geq0}{\max} \ \ee_{\cir}(V,r).\label{eq:circular}
\end{align}

To solve problem \eqref{eq:circular}, it is first noted that the numerator of $\ee_{\cir}(V,r)$ in \eqref{eq:eeCir} is independent of the UAV speed $V$. Thus, for any fixed radius $r$, the optimal speed $V$ is obtained by minimizing the denominator of \eqref{eq:eeCir}, which can be readily obtained as
\begin{align}
V_{\cir}^\star(r)=\left(\frac{c_2}{3(c_1+c_2/(g^2r^2))}\right)^{1/4}.
\end{align}
The corresponding UAV power consumption, i.e., the denominator of \eqref{eq:eeCir}, reduces to a univariate function of $r$ as
\begin{align}
P_\cir^\star(r)=\left(3^{-3/4}+3^{1/4} \right)c_2^{3/4}\left(c_1+\frac{c_2}{g^2r^2}\right)^{1/4}.
\end{align}
Thus, by discarding the constant terms and defining $z=r^2$, problem \eqref{eq:circular} reduces to a univariate optimization problem given by
\begin{align}
\underset{z\geq 0}{\max} \ \eta(z)\triangleq \frac{\ln\left(1+\frac{\gamma_0}{H^2+z}\right)}{\left(c_1+\frac{c_2}{g^2z}\right)^{1/4}}.\label{eq:findZ}
\end{align}
Since $\eta(0)=0$ and $\lim_{z\rightarrow \infty} \eta(z) \rightarrow 0$, there must exist a finite optimal solution $z^\star$ to problem \eqref{eq:findZ}, which can be efficiently found numerically. Furthermore, in the low-SNR regime with $\gamma_0\ll H^2$, by applying the result $\ln(1+x)\approx x$ if $x\ll 1$, the optimal solution to \eqref{eq:findZ} can be obtained in closed-form as
\begin{align}
z^\star= \frac{3c_2}{8c_1g^2}\left[\Big(1+\frac{16H^2c_1g^2}{9c_2}\Big)^{1/2}-1 \right].\label{eq:zLowSNR}
\end{align}

%


As the optimal value to problem \eqref{eq:findZ} is guaranteed to be positive, a strictly positive energy efficiency is always achievable by the optimized circular trajectory for any $T>0$, which demonstrates its superiority over the rate-maximization or energy-minimization designs considered in Section~\ref{sec:unConstrained}.

   \section{Energy Efficiency Maximization with Generally Constrained Trajectory}\label{sec:EfficientSol}
   The preceding two sections study the energy efficiency for unconstrained trajectory designs. In practice, the UAV's trajectory $\q(t)$ may need to satisfy a number of practical constraints, such as those on its initial/final states (including location and velocity) and the maximum speed/acceleration. They are mathematically represented as
\begin{align}
\C_1:\ & \q(0)=\q_0 \\
\C_2: \ & \q(T)=\q_F \\
\C_3:\ & \|\mathbf v(t)\|\leq V_{\max}, \forall t\\
\C_4:\ & \mathbf v(0)=\mathbf v_0\\
\C_5:\ & \mathbf v(T)=\mathbf v_F\\
\C_6:\ & \|\mathbf a(t)\|\leq a_{\max}, \forall t,
\end{align}
where $\mathbf q_0, \mathbf q_F\in \mathbb{R}^{2\times 1}$ denote the UAV's destined  initial and final locations, respectively, $\vbd_0, \vbd_F\in \mathbb{R}^{2\times 1}$ are the desired initial and final velocities, respectively, and $V_{\max}$ and $a_{\max}$ represent the maximum speed and acceleration, respectively. In this case, the energy-efficiency maximization problem can be formulated as
   \begin{align}
   \mathrm{(P1):}\ \underset{\q(t)}{\max} \  & \ee(\q(t)) \notag \\
   \text{s.t.} &  \ \C_1 \text{--} \C_6, \notag
   \end{align}
   where $\ee(\q(t))$ is given in \eqref{eq:EE}. Problem (P1) is  difficult to be directly solved for two reasons. Firstly, it requires the optimization of the continuous function $\q(t)$, as well as its first- and second-order derivatives $\mathbf v(t)$ and $\mathbf a(t)$,  which essentially involves an infinite number of optimization variables. Secondly, the objective function in $\ee(\q(t))$ is given by the fraction of two integrals, which both lack closed-form expressions. In the following, an efficient algorithm is proposed for (P1) based on two main techniques: discrete linear state-space approximation and sequential convex optimization.


To this end, it is first noted that the energy consumption in \eqref{eq:ER} can be upper-bounded by
\begin{align}
\hspace{-1ex}\bar E(\q(t))\leq  & \int_{0}^T \left[c_1\|\mathbf v(t)\|^3+\frac{c_2}{\|\mathbf v(t)\|}\bigg(1+\frac{\|\mathbf a(t)\|^2}{g^2}\bigg)\right]dt\notag \\
&+\Delta_K\triangleq \bar E_{\mathrm{ub}}(\q(t)),\label{eq:ERUB}
\end{align}
with $\Delta_K \triangleq \frac{1}{2}m \left(\|\mathbf v(T)\|^2-\|\mathbf v(0)\|^2\right)$ denoting the change of the UAV's kinetic energy, which is fixed with the initial and final velocity constraints $\C_4$ and $\C_5$. 
Note that the upper bound in \eqref{eq:ERUB} is tight for constant-speed flight, in which case $\mathbf a(t)^T \mathbf v(t)=0$, $\forall t$. Therefore, the  energy efficiency in \eqref{eq:EE} is lower-bounded by
\begin{align}
\ee(\q(t)) & \geq  \ee_{\mathrm{lb}}(\q(t))\triangleq \frac{\bar R(\q(t))}{\bar E_{\mathrm{ub}}(\q(t))}\notag \\
& =\frac{B \int_0^T \log_2\left(1+\frac{\gamma_0}{H^2+\|\q(t)\|^2}\right) dt}{\int_{0}^T \left[c_1\|\mathbf v(t)\|^3+\frac{c_2}{\|\mathbf v(t)\|}\bigg(1+\frac{\|\mathbf a(t)\|^2}{g^2}\bigg)\right]dt+\Delta_K}.
\end{align}
Thus, (P1) can be approximately solved by maximizing its lower bound as
\begin{align}
   \mathrm{(P1'):} & \ \underset{\q(t)}{\max} \  \ee_{\mathrm{lb}}(\q(t))   \notag \\
   \text{s.t.} &  \ \C_1 \text{--} \C_6.
   \end{align}

   To obtain a more tractable optimization problem, we apply the discrete linear state-space approximation to $\mathrm{(P1')}$. Since $\vbd(t)\triangleq \dot{\q}(t)$ and $\abd(t)\triangleq \dot{\vbd}(t)$ are respectively the time-varying velocity and acceleration vectors associated with the UAV trajectory $\q(t)$, for any infinitesimal time step $\delta_t$, we have the following results based on the first- and second-order Taylor approximations,
\begin{align}
\vbd(t+\delta_t)&\approx \vbd(t)+\abd(t) \delta_t, \ \forall t, \label{eq:vTaylor} \\
\q(t+\delta_t)& \approx \q(t)+\vbd(t)\delta_t+\frac{1}{2}\abd(t)\delta_t^2, \ \forall t.\label{eq:qTaylor}
\end{align}

As a result, by discretizing the time horizon $T$ into $N+2$ slots with step size $\delta_t$, i.e., $t=n\delta_t$, $n=0,1,\cdots,N+1$, the UAV's trajectory $\q(t)$ can be well characterized by the discrete-time UAV location $\q[n]\triangleq \q(n\delta_t)$, the velocity $\vbd[n]\triangleq\vbd(n\delta_t)$, as well as the acceleration $\abd[n]\triangleq \abd(n\delta_t)$, $n=0,1,\cdots, N+1$. As a result, we have the following  discrete state-space model based on \eqref{eq:vTaylor} and \eqref{eq:qTaylor},
\begin{align}
\vbd[n+1]&= \vbd[n]+\mathbf a[n] \delta_t, \\
\q[n+1]&= \q[n]+\vbd[n] \delta_t+\frac{1}{2}\mathbf a[n]\delta_t^2, \ n=0,1,\cdots N,
\end{align}
which is linear with respect to $\q[n], \mathbf v[n]$ and $\mathbf a[n]$.
 As a result, problem $\mathrm{(P1')}$ can be rewritten as
\begin{align}
  \mathrm{(P2)}: & \underset{\substack{\{\q[n], \mathbf v[n]\\ \mathbf a[n]\}}}{\max}  \frac{B \sum_{n=1}^N \log_2\left(1+\frac{\gamma_0}{H^2+\|\q[n]\|^2}\right)}{\sum_{n=1}^{N} \left(c_1 \|\mathbf v[n]\|^3+\frac{c_2}{\|\mathbf v[n]\|}\left(1+\frac{\|\mathbf a[n]\|^2}{g^2} \right)\right)+\frac{\Delta_K}{\delta_t}}\notag \\
\text{s.t.} \ & \q[n+1]= \q[n]+\vbd[n] \delta_t+\frac{1}{2}\mathbf a[n]\delta_t^2, \ n=0,\cdots N \label{eq:P21stConstr} \\
& \vbd[n+1]= \vbd[n]+\mathbf a[n] \delta_t, \ n=0,\cdots N \\
& \q[0]=\q_0,\ \q[N+1]=\q_F \label{eq:qDiscrete}\\
& \mathbf v[0]=\mathbf v_0,\ \mathbf v[N+1]=\mathbf v_F \\
& \|\mathbf v[n]\| \leq V_{\max}, \ n=1,\cdots, N \\
& \|\mathbf a[n]\|\leq a_{\max}, \ n=0,\cdots N,\label{eq:P2LastConstr}
\end{align}
where \eqref{eq:qDiscrete}--\eqref{eq:P2LastConstr} represent the discrete equivalents of $\C_1 \text{--} \C_6$.

Note that all constraints of problem $\mathrm{(P2)}$ are convex. However, the objective is a fractional function with  a non-concave numerator over a non-convex denominator, and hence $\mathrm{(P2)}$  is neither a convex nor quasi-convex problem, which thus cannot be directly solved with the standard convex optimization techniques. Fortunately, by applying the sequential convex optimization technique, an efficient solution can be obtained which is guaranteed to satisfy the Karush-Kuhn-Tucker (KKT) conditions of $\mathrm{(P2)}$. This implies that at least a local optimal solution can be found for the problem. To this end, we first reformulate $\mathrm{(P2)}$ by introducing slack variables $\{\tau_n\}$ as
\begin{align}
  \mathrm{(P2.1)}: & \underset{\substack{\{\q[n], \mathbf v[n]\\ \mathbf a[n],\tau_n\}}}{\max} \frac{B \sum_{n=1}^N \log_2\left(1+\frac{\gamma_0}{H^2+\|\q[n]\|^2}\right)}{\sum_{n=1}^{N} \left(c_1 \|\mathbf v[n]\|^3+\frac{c_2}{\tau_n}+\frac{c_2\|\mathbf a[n]\|^2}{g^2 \tau_n} \right)+\frac{\Delta_K}{\delta_t}}\notag \\
\text{s.t.} &\ \eqref{eq:P21stConstr}\text{--} \eqref{eq:P2LastConstr}, \notag\\
& \  \tau_n\geq 0,\ \forall n, \\
& \|\mathbf v[n]\|^2 \geq \tau_n^2,\  \forall n. \label{eq:nonCvxContr0}
 \end{align}

 It can be shown that at the optimal solution to $\mathrm{(P2.1)}$, we must have $\tau_n=\|\mathbf v[n]\|$, $\forall n$, since otherwise one can always increase $\tau_n$ to obtain a strictly larger objective value. Thus, $\mathrm{(P2.1)}$ is equivalent to $\mathrm{(P2)}$. With such a reformulation, the denominator of the objective function in $\mathrm{(P2.1)}$ is now jointly convex with respect to $\{\mathbf v[n], \mathbf a[n], \tau_n\}$, but with the new non-convex constraint \eqref{eq:nonCvxContr0}.
 To tackle this non-convex constraint, a local convex approximation is applied. Specifically, since $\|\mathbf v[n]\|^2$ is a convex and differentiable function with respect to $\mathbf v[n]$, for any given local point $\{\mathbf v_j[n]\}$, we have
 \begin{align}
 \|\mathbf v[n]\|^2 & \geq \|\mathbf v_j[n]\|^2 + 2 \mathbf v_j^T[n]\left( \mathbf v[n]-\mathbf v_j[n]\right)\notag \\
 &\triangleq \psi_{\mathrm{lb}}(\mathbf v[n]), \ \forall \mathbf v[n],\label{eq:vnLB}
 \end{align}
 where the equality holds at the point $\mathbf v[n]=\mathbf v_j[n]$. Note that \eqref{eq:vnLB} follows from the fact that the first-order Taylor expansion of a convex differentiable function is its global under-estimator \cite{202}. Furthermore, at the local point $\mathbf v_j[n]$, both the function $\|\mathbf v[n]\|^2$ and its lower bound $\psi_{\mathrm{lb}}(\mathbf v[n])$ have the identical gradient, which is equal to $2\mathbf v_j[n]$.

 Define the new constraint,
 \begin{align}
 \psi_{\mathrm{lb}}(\mathbf v[n])\geq \tau_n^2, \ \forall n, \label{eq:newConstrvn}
 \end{align}
 which is convex since $\psi_{\mathrm{lb}}(\mathbf v[n])$ is linear with respect to $\mathbf v[n]$. Then the inequality in \eqref{eq:vnLB} shows that the convex constraint \eqref{eq:newConstrvn} always implies the non-convex constraint  \eqref{eq:nonCvxContr0}, but the reverse is not true in general.

 Similarly, to tackle the non-concavity of the numerator of the objective function in $\mathrm{(P2.1)}$, for any local point $\{\q_j[n]\}$, define the function
\begin{align}
\bar R_{\lb}\left(\{\q[n]\}\right)=B\sum_{n=1}^N \Big[\alpha_j[n]-\beta_j[n]\left(\|\q[n]\|^2-\|\q_j[n]\|^2 \right)\Big],\label{eq:Rlb}
\end{align}
where
\begin{align}
&\alpha_j[n]=\log_2\left(1+\frac{\gamma_0}{H^2+\|\q_j[n]\|^2}\right),\\
&\beta_j[n]=\frac{(\log_2 e) \gamma_0}{ (H^2+\gamma_0+\|\q_j[n]\|^2)(H^2+\|\q_j[n]\|^2)}, \ \forall n.
\end{align}
Note that $\bar R_{\lb}\left(\{\q[n]\}\right)$ is a concave function with respect to $\{\q[n]\}$. Furthermore, we have the following result.
\begin{theorem}\label{theo:LB}
For any given $\{\q_j[n]\}$, we have
\begin{align}
\bar R(\{\q[n]\})& \triangleq B \sum_{n=1}^N \log_2\left(1+\frac{\gamma_0}{H^2+\|\q[n]\|^2}\right)\notag \\
&\geq \bar R_{\lb}(\{\q[n]\}), \ \forall \q[n],\label{eq:LB}
\end{align}
where the equality holds at the point $\q[n]=\q_j[n]$, $\forall n$. Furthermore, at the point $\q[n]=\q_j[n]$, $\forall n$, both $\bar R(\{\q[n]\})$ and $R_{\lb}(\{\q[n]\})$ have identical gradient, i.e.,  $\nabla \bar R=\nabla \bar R_{\lb}$.
\end{theorem}
\begin{IEEEproof}
Please refer to Appendix~\ref{A:LB}.
\end{IEEEproof}

 As a result, for any given local point $\{\q_j[n], \mathbf v_j[n]\}$, define the following optimization problem,
\begin{align}
  \mathrm{(P2.2)}: & \underset{\substack{\{\q[n], \mathbf v[n]\\ \mathbf a[n],\tau_n\}}}{\max} \frac{\bar R_{\lb}(\{\q[n]\})}{\sum_{n=1}^{N} \left(c_1 \|\mathbf v[n]\|^3+\frac{c_2}{\tau_n}+\frac{c_2\|\mathbf a[n]\|^2}{g^2 \tau_n} \right)+\frac{\Delta_K}{\delta_t}}\notag \\
\text{s.t.} &\ \eqref{eq:P21stConstr}\text{--} \eqref{eq:P2LastConstr}, \notag\\
& \ \tau_n\geq 0, \forall n, \notag\\
&  \psi_{\mathrm{lb}}(\mathbf v[n])\geq \tau_n^2, \forall n.\notag
 \end{align}
 Based on the previous discussions, it readily follows that the objective value of $\mathrm{(P2.2)}$ gives a lower bound to that of problem $\mathrm{(P2.1)}$. Furthermore, problem $\mathrm{(P2.2)}$ is a fractional maximization problem, with a concave numerator and a convex denominator, as well as all convex constraints. Thus, $\mathrm{(P2.2)}$ can be efficiently solved via the bisection method \cite{202} or the standard Dinkelbach's algorithm for fractional programming \cite{802}.

 Thus, the  original non-convex problem $\mathrm{(P2.1)}$  can be solved by iteratively  optimizing $\mathrm{(P2.2)}$ with the local point $\{\mathbf q_j[n], \mathbf v_j[n]\}$ updated in each iteration, which is summarized in Algorithm~\ref{Algo:succ}.
 \begin{algorithm}[H]
\caption{Sequential convex optimization for $\mathrm{(P2.1)}$.}\label{Algo:succ}
\begin{algorithmic}[1]
\STATE Initialize $\{\q_0[n],\mathbf v_0[n]\}$. Let $j=0$.
\REPEAT
\STATE Solve problem $\mathrm{(P2.2)}$ for the given local point $\{\q_j[n], \mathbf v_j[n]\}$, and denote the optimal solution as $\{\q_j^*[n], \mathbf v_j^*[n]\}$.
\STATE Update the local point $\q_{j+1}[n]=\q_j^*[n]$ and  $\mathbf v_{j+1}[n]=\mathbf v_j^*[n]$, $\forall n$.
\STATE Update $j=j+1$.
\UNTIL{Converges to a prescribed accuracy}.
\end{algorithmic}
\end{algorithm}

Let $\ee_{\mathrm{lb},j}^*$ denote the corresponding optimal value of $\mathrm{(P2.1)}$ obtained via Algorithm~\ref{Algo:succ} at the $j$th iteration. We have the following result.
\begin{lemma}\label{lemma:KKTOpt}
The energy efficiency lower bound $\ee_{\mathrm{lb},j}^*$ obtained in  Algorithm~\ref{Algo:succ} is monotonically non-decreasing, i.e., $\ee_{\mathrm{lb},j}^*\geq \ee_{\mathrm{lb},j-1}^*$, $\forall j\geq 1$. Furthermore, the sequence $\{\q_j^*[n], \mathbf v_j^*[n]\}$, $j=0,1,\cdots$, converges to a point fulfilling the KKT optimality conditions of the original non-convex problem $\mathrm{(P2.1)}$.
\end{lemma}
\begin{IEEEproof}
Lemma~\ref{lemma:KKTOpt} directly follows from Proposition 3 of reference \cite{768}, by making use of Theorem~\ref{theo:LB} as well as the fact that the lower bound $\psi_{\mathrm{lb}}(\mathbf v[n])$ has both identical value and identical gradient as $\|\mathbf v[n]\|^2$ at the local point $\mathbf v[n]=\mathbf v_j[n]$. The details are omitted for brevity.
\end{IEEEproof}

As a final remark, it is noted that Algorithm~\ref{Algo:succ} can also be applied for energy efficiency maximization with unconstrained trajectory optimization by discarding $\C_1 \text{--} \C_6$, which gives an alternative energy-efficient design without restricting to circular trajectory considered in Section~\ref{sec:Circular}. Furthermore, Algorithm~\ref{Algo:succ} can be similarly applied for the rate-maximization and energy-minimization designs for constrained trajectories, since they correspond to the special cases of $\mathrm{(P2)}$ by only maximizing/minimizing the numerator and denominator, respectively. These special cases  will be considered in the numerical simulations given in the next section.

\section{Numerical Results}\label{sec:Numerical}
In this section, numerical results are provided to validate the proposed design. The UAV altitude is fixed at $H=100$m. The communication bandwidth is $B=1$MHz and the noise power spectrum density at the GT receiver is assumed to be $N_0=-170$dBm/Hz. Thus, the corresponding noise power is $\sigma^2=N_0B=-110$dBm. We assume that the UAV transmission power is $P=10$dBm (or 0.01W), and the reference channel power is $\beta_0=-50$dB.   As a result, the maximum SNR achieved when the UAV is just above the GT can be obtained as $30$dB. Furthermore, we assume that $c_1=9.26\times 10^{-4}$ and $c_2=2250$, such that the UAV's energy-minimum speed is $V_{\emi}=30$m/s and the corresponding minimum propulsion power consumption is $P_{\emi}=100$W. Note that we have $P\ll P_{\emi}$, thus the UAV transmission power can be practically ignored.

\begin{figure}
\centering
\includegraphics[scale=0.6]{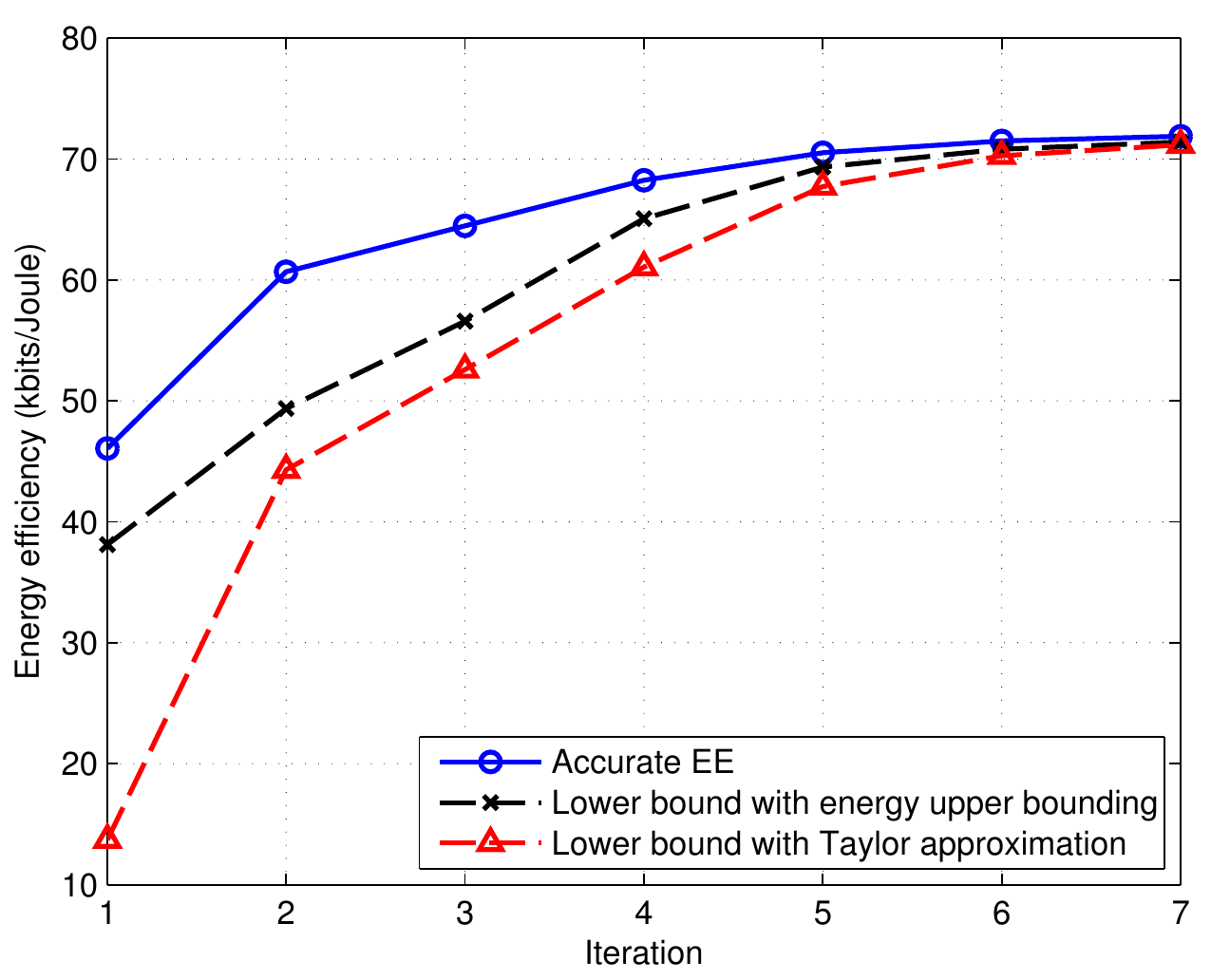}
\caption{Convergence of Algorithm~\ref{Algo:succ}.}\label{F:Convergence}
\end{figure}

\begin{table*}
\centering
\caption{Performance comparison for various designs with unconstrained trajectory optimization. 
}\label{table:circularVSSCO}
\begin{tabular}{|p{3cm}|p{1.5cm}|p{1.5cm}|p{1.5cm}|p{1.5cm}|p{1.5cm}|}
\hline
& {Average speed (m/s)} & {Average acceleration (m$^2$/s)} & {Average rate (Mbps)} & {Average power (Watts)} & {Energy efficiency (kbits/Joule)} \\
\hline
{Rate-maximization} & $0$ & $0$ & $9.97$ & $\infty$ & $0$\\
\hline
{Energy-minimization} & $30$ & $0$ &$6.06$ & $100$ & $60.6$ \\
\hline
{EE-maximization, circular} & $25.20$ & $4.02$ & $8.16$ & $119.10$ & $68.56$ \\
\hline
{EE-maximization, sequentially optimized} & $25.67$ & $3.24$& $8.34$ & $116.02$ & $71.89$ \\
\hline
\end{tabular}
\end{table*}

\subsection{Unconstrained Trajectory Optimization}
We first consider the case of unconstrained trajectory optimization in the absence of $\C_1\text{--}\C_6$. Besides the three specific trajectory designs considered in Section~\ref{sec:unConstrained} and Section~\ref{sec:Circular}, namely the rate-maximization, energy-minimization, and energy-efficient circular designs,  the energy-efficiency maximization design by applying the similar sequential convex optimization proposed in Algorithm~\ref{Algo:succ} but without trajectory constraints is also included for comparison.

Under this setup, Fig.~\ref{F:Convergence} shows the convergence of Algorithm~\ref{Algo:succ} with randomly generated initial points. Note that  Fig.~\ref{F:Convergence} consists of three curves: the ``Accurate EE'' corresponds to the exact energy efficiency based on the energy consumption model \eqref{eq:ER}, the ``Lower bound with energy upper bounding'' is that due to the energy upper bounding given in \eqref{eq:ERUB}, and the ``Lower bound with Taylor approximation'' corresponds to the optimal value of $\mathrm{(P2.2)}$ by using local convex approximation. Fig.~\ref{F:Convergence} shows that Algorithm~\ref{Algo:succ} monotonically converges, as expected from Lemma~\ref{lemma:KKTOpt}. Furthermore, it also shows that the adopted lower bounds for the purpose of efficient convex optimization are rather tight, especially as the algorithm converges.

\begin{figure}
\centering
\includegraphics[scale=0.6]{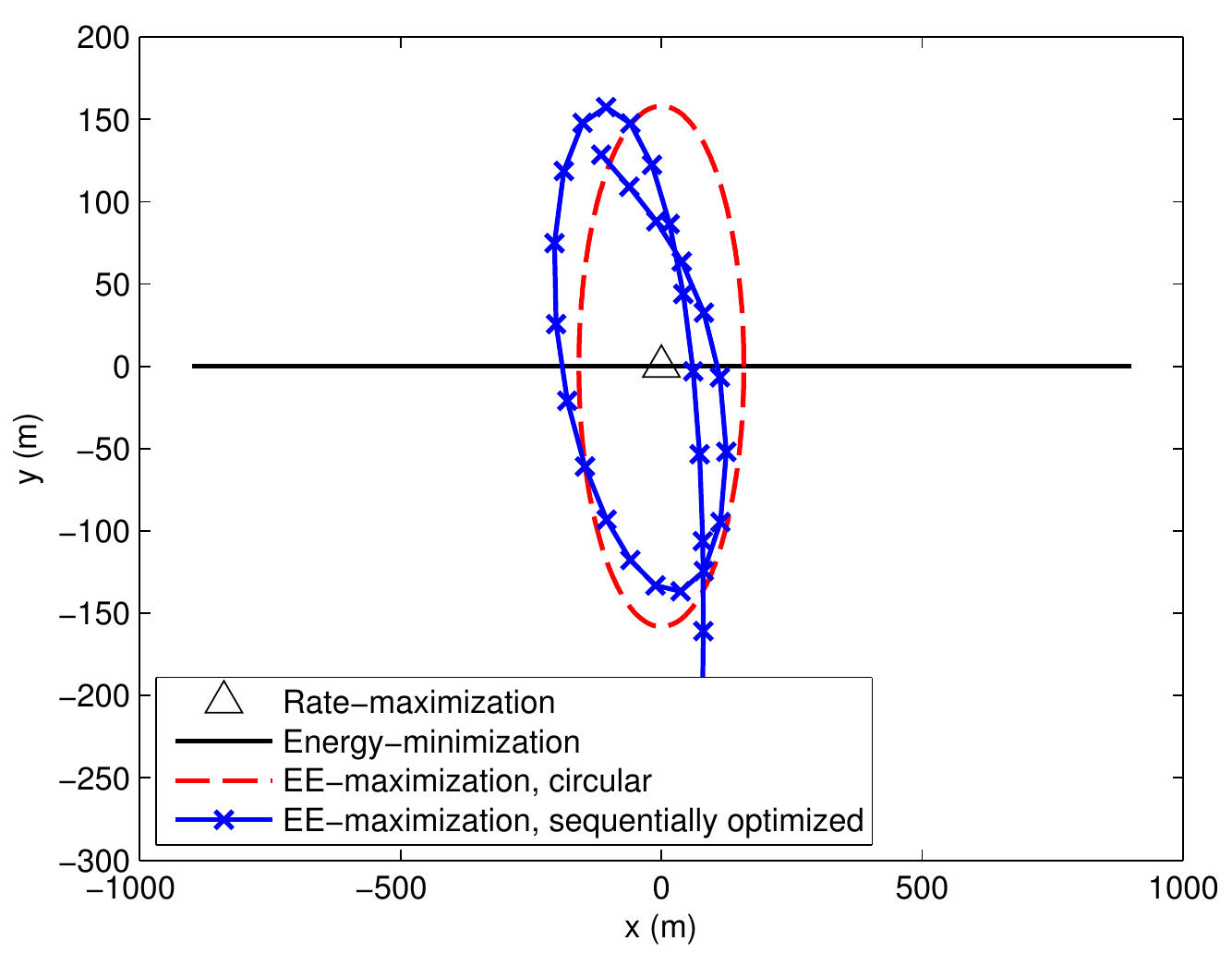}
\caption{Trajectory comparison for unconstrained optimization.}\label{F:TrajectoryCircularVSSCO}
\end{figure}

 Fig.~\ref{F:TrajectoryCircularVSSCO} shows the obtained UAV trajectories projected onto the horizontal plane for four different designs with  $T=60$s. As shown in Section~\ref{sec:unConstrained}, the unconstrained rate-maximization and energy-minimization trajectories correspond to stationary hovering with $V=0$ and steady straight flight with power-minimum speed $V_{\emi}=30$m/s, respectively. On the other hand, for the EE-maximization circular trajectory, the optimal flight radius and speed are $r^\star=158$m and $V^\star=25.67$m/s, respectively. It is observed that the converged trajectory  via applying the sequential convex optimization for EE maximization is rather similar to the circular trajectory. 
 Table~\ref{table:circularVSSCO} gives a detailed comparison for these four different designs in terms of the average UAV speed, acceleration, achievable communication rate, power consumption, and the overall energy efficiency. It is found that for the unconstrained trajectory setup, the simple circular trajectory with optimized flight radius and speed gives a comparable performance as that obtained via sequential convex optimization, and they both notably outperform the rate-maximization or energy-minimization designs in terms of energy efficiency.

\subsection{Constrained Trajectory Optimization}
\begin{figure}
\centering
\includegraphics[scale=0.6]{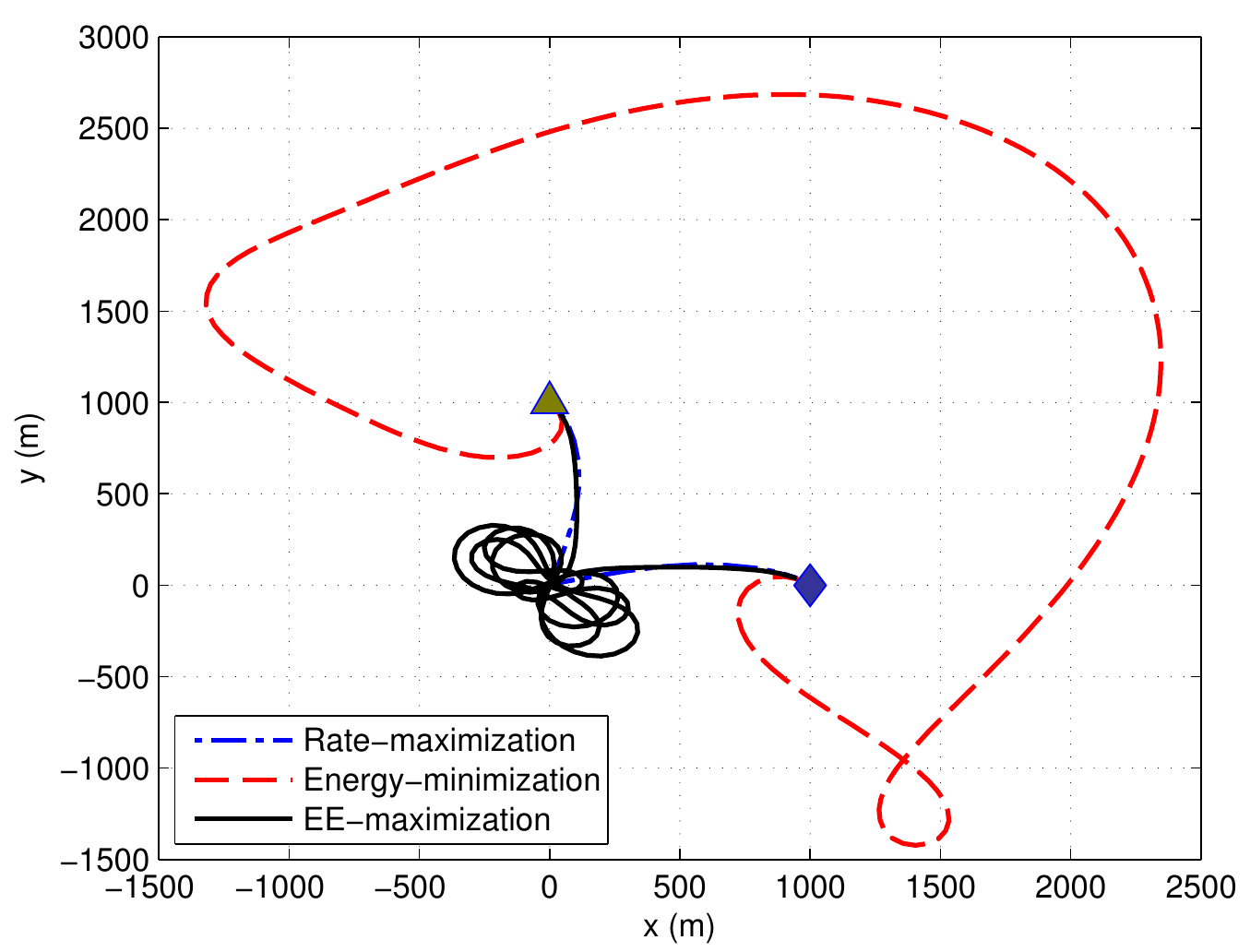}
\caption{Trajectory comparison with constrained optimization. The triangle and diamond denote the initial and final UAV locations, respectively.\vspace{-3ex}}\label{F:TrajectoryConstrainedOpt}
\end{figure}

Next, we consider the constrained trajectory optimization as studied in Section~\ref{sec:EfficientSol}. As shown in Fig.~\ref{F:TrajectoryConstrainedOpt}, we let $\q_0= [1000, \ 0]^T, \q_F=[0,\ 1000]^T, \mathbf v_0=\mathbf v_F=30 \vec{\mathbf v}_{0F}$, with $\vec{\mathbf v}_{0F}\triangleq (\q_F-\q_0)/\|\q_F-\q_0\|$ denoting the direction from $\q_0$ to $\q_F$. Furthermore, the maximum UAV speed and acceleration are  set to $V_{\max}=100$m/s and $a_{\max}=5$m/s$^2$, respectively, and the operation time is  $T=400$s. For the sequential convex optimization in Algorithm~\ref{Algo:succ}, the initial points are set to be the direct path from $\q_0$ to $\q_F$. 

Fig.~\ref{F:TrajectoryConstrainedOpt} shows the obtained trajectories with the rate-maximization, energy-minimization, and EE-maximization designs. Note that the rate-maximization and energy-minimization trajectories are obtained with the similar sequential convex optimization technique as in Algorithm~\ref{Algo:succ} with the objective function modified accordingly. It is observed from Fig.~\ref{F:TrajectoryConstrainedOpt} that besides satisfying the initial and final location constraints, all the three trajectories are tangential to the direction vector $\vec{\mathbf v}_{0F}$ at both the initial and final locations $\q_0$ and $\q_F$, due to the imposed initial and final velocity constraints. Furthermore, for the rate-maximization trajectory, it is found that the UAV hovers around the GT (i.e., the origin) for the maximum possible duration to maintain the best communication channel. In contrast, the energy-minimization design yields a trajectory mostly with a large turning radius since it is in general less power-consuming. However, such a trajectory is expected to incur rather poor communication channels due to the large distance from the GT. On the other hand, with the proposed EE-maximization trajectory design, upon approaching the GT, the UAV hovers around following an approximately ``8''-shape path, which is expected to maintain a sufficiently good communication channel yet without excessive power consumption. The performance comparison of the above three trajectory designs is given in Table~\ref{table:Constrained}. It is observed that the proposed EE-maximization design achieves a good balance between rate maximization (column 4) and power minimization (column 5), and hence achieves significantly higher energy efficiency (column 6) than the rate-maximization and energy-minimization designs.


\begin{table*}
\centering
\caption{Performance comparison for various designs with constrained trajectory optimization. 
}\label{table:Constrained}
\begin{tabular}{|p{3cm}|p{1.5cm}|p{1.5cm}|p{1.5cm}|p{1.5cm}|p{1.5cm}|}
\hline
& {Average speed (m/s)} & {Average acceleration (m$^2$/s)} & {Average rate (Mbps)} & {Average power (Watts)} & {Energy efficiency (kbits/Joule)} \\
\hline
{Rate-maximization} & $8.40$ & $3.74$ & $9.56$ & $585.66$ & $16.32$\\
\hline
{Energy-minimization} & $29.61$ & $1.10$ & $2.01$ & $102.74$ & $19.60$ \\
\hline
{EE-maximization} & $26.03$ & $3.21$ & $7.48$ & $118.57$ & $63.08$ \\
\hline
\end{tabular}
\end{table*}


%


\section{Conclusion}\label{sec:Conclusion}
This paper studies the energy-efficient UAV communication via trajectory optimization by taking into account the propulsion energy consumption of the UAV. A theoretical model on the UAV's propulsion energy consumption is derived, based on which the energy efficiency of UAV communication is defined. For the case of unconstrained trajectory optimization, it is shown that both the rate-maximization and energy-minimization designs lead to vanishing energy efficiency and thus are energy-inefficient in general. We then consider a practical circular trajectory with optimized flight radius and speed for maximizing energy efficiency. Furthermore, for the generally constrained trajectory optimization, an efficient algorithm is proposed to maximize the energy efficiency based on linear state-space approximation and sequential convex optimization techniques. Numerical results show that the proposed designs achieve significantly higher energy efficiency than heuristic rate-maximization and energy-minimization designs for UAV communications.

\appendices
\section{UAV Propulsion Energy Consumption Model}\label{A:energyModel}
In this appendix, we derive the propulsion energy consumption model of fixed-wing UAVs.

\begin{figure}
\centering
\includegraphics[scale=0.6]{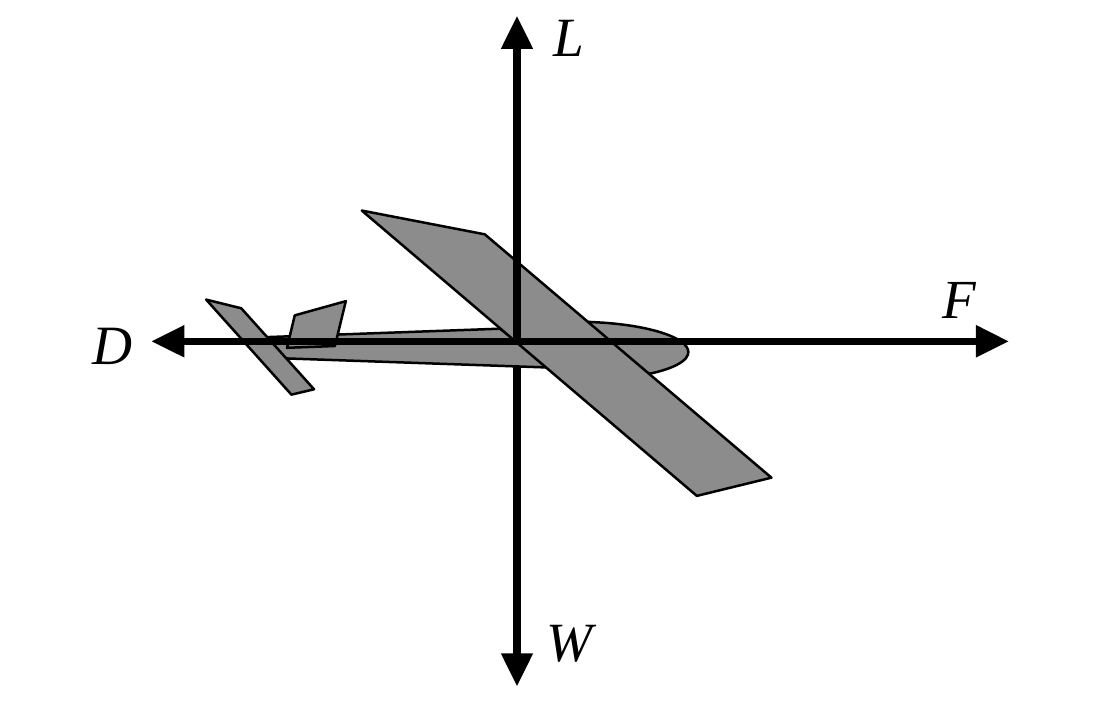}
\caption{A schematic of the forces on an aircraft in straight-and-level flight.}\label{F:ForcesUAV}
\end{figure}

\subsubsection{Forces on the Aircraft}
As shown in Fig.~\ref{F:ForcesUAV}, an aircraft aloft is in general subject to four forces:  weight, drag,  lift, and thrust \cite{766}.
\begin{itemize}
\item {\it Weight} ($W$): the force of gravity, $W=mg$, with $m$ denoting the aircraft's mass including all its payload, and $g$ is the gravitational acceleration in m/s$^2$.

\item {\it Drag} ($D$): the aerodynamic force component parallel to the airflow direction. For zero wind speed, $D$ is in the opposite direction of the aircraft's motion. 

\item {\it Lift} ($L$): the aerodynamic force component normal to the drag and pointing upward.

\item {\it Thrust} ($F$): the force produced by the aircraft engine, which overcomes the drag to move the aircraft forward.
\end{itemize}

For an aircraft moving at a subsonic speed $V$ (as usually the case), the drag can be expressed as \cite{766}
\begin{align}
D=\frac{1}{2}\rho C_{D_0} S V^2+ \frac{2L^2}{(\pi e_0 \mathcal{A_R})  \rho S V^2}, \label{eq:Drag}
\end{align}
where $\rho$ is the air density in kg/m$^3$, $C_{D_0}$ is the zero-lift drag coefficient, 
$S$ is a reference area (e.g., the wing area), $e_0$ is the Oswald efficiency (or wing span efficiency) with typical value between $0.7$ and
$0.85$ \cite{766}, $\mathcal{A_R}$ is the aspect ratio of the wing, i.e., the ratio of the wing span to its aerodynamic breadth. The first term in \eqref{eq:Drag} is known as the {\it parasitic drag}, which is a combination of multiple drag components such as form drag, skin friction drag and interference drag, etc. The parasitic drag increases quadratically with the vehicle speed $V$. The second term in \eqref{eq:Drag} is called the {\it lift-induced drag}, which is the resulting drag force due to wings redirecting air to generate lift $L$. 

The drag expression \eqref{eq:Drag} can be rewritten as
\begin{align}
D=c_1 V^2+ \frac{c_2\kappa^2}{V^2},\label{eq:Drag2}
\end{align}
where
\begin{align}
c_1\triangleq \frac{1}{2}\rho C_{D_0} S,  \ c_2\triangleq \frac{2W^2}{(\pi e_0 \mathcal{A_R})  \rho S}\label{eq:c1c2}
\end{align}
are two constant parameters, and
$\kappa\triangleq L/W$
is known as the {\it load factor}, i.e., the ratio of the aircraft's lift to its weight. Note that to maintain a level flight (i.e., a flight with constant altitude), we must have $\kappa \geq 1$ since the lift must at least balance the aircraft weight to avoid descending.
  It is then not difficult to conclude from \eqref{eq:Drag2} that $\forall V> 0$, we have $D\geq D_{\min}$, where
$D_{\min}=2\sqrt{c_1 c_2}$
is the minimum drag incurred, which corresponds to  $\kappa=1$ and the drag-minimum speed $V_{\dm}=(c_2/c_1)^{1/4}$.
%
%
%
%
%

\subsubsection{Power Required for Straight-and-Level Flight}
{\it Straight-and-level} flight refers to the flight in which a constant  heading (or direction) and altitude are maintained. This implies that: (i) the horizontal acceleration, if any, must be in parallel with the aircraft's flying direction so that no turning occurs; (ii) the lift and weight are balanced so that there is no vertical acceleration. We thus have the following equations (as illustrated in Fig.~\ref{F:ForcesUAV}):
\begin{align}
 L=W, \quad  F-D=ma,\label{eq:Newton2ndLaw}
\end{align}
where $a$ is the acceleration along the aircraft's flying direction, i.e., $a>0$ for accelerating and $a<0$ for decelerating. 
When $a\geq -D/m$, 
we have $F\geq 0$. In this case, the thrust $F$ generated by the engine is in the same direction as the aircraft's motion, or {\it forward thrust}. In contrast, if $a<-D/m$, we have $F<0$, in which case the thrust $F$ generated by the engine must be in the opposite direction as the aircraft's motion, or {\it reverse thrust}. One sufficient (but not necessary) condition to ensure forward thrust operation is $a\geq -D_{\min}/m$.

By considering an infinitesimal time interval such that the speed can be regarded as unchanged, it follows from \eqref{eq:Drag2} and \eqref{eq:Newton2ndLaw} that the power required for straight-and-level flight with speed $V$ and acceleration $a$ can be expressed as
\begin{align}
P_{\req}(V,a)&=|F|V
=\left | c_1 V^3+ \frac{c_2}{V}+maV \right| ,\label{eq:Preq}
\end{align}
where we have used the fact that power is equal to force times speed. Note that the magnitude operator in \eqref{eq:Preq} is necessary for the unusual case of reverse thrust with $F<0$ when the aircraft needs to have an abrupt deceleration.


\subsubsection{Power Required for Banked Level Turn}
\begin{figure}
\centering
\includegraphics[scale=0.7]{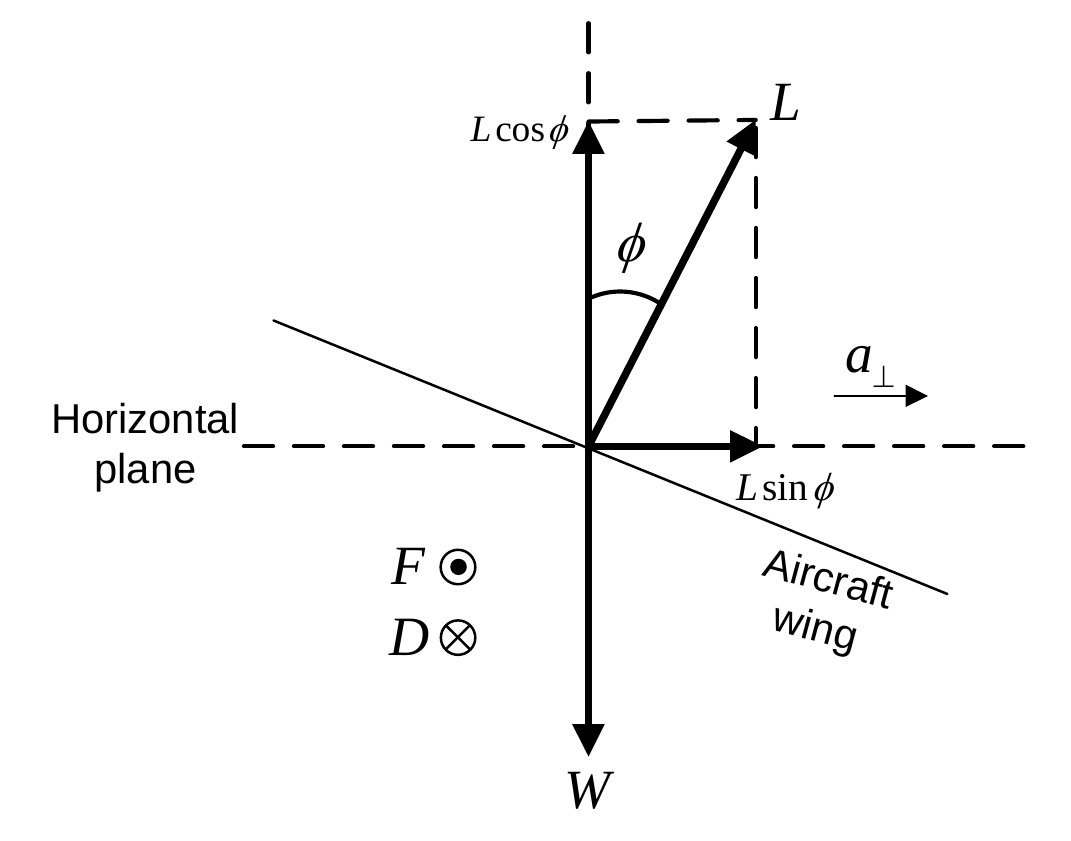}
\caption{A schematic of the forces on an aircraft in banked level turn. The aircraft moves normal to the page. The horizontal component of the lift $L\sin\phi$ causes the centrifugal acceleration $a_{\perp}$ to change the heading.}\label{F:BankedLevelTurn}
\end{figure}

For a fixed-wing aircraft to change its flying direction, the aircraft must roll to a banked position so that the lift $L$ generates a lateral (horizontal) component to support the centrifugal acceleration, i.e., the acceleration component normal to the velocity. As illustrated in Fig.~\ref{F:BankedLevelTurn}, let $\phi$ denote the bank angle, i.e., the angle between the vertical plane and the aircraft's symmetry plane. We then have the following equations,
\begin{align}
& L\cos \phi = W, \quad  L\sin \phi = m a_{\perp},\label{eq:L3}\\
& F-D=m a_{\parallel} \label{eq:F1},
\end{align}
where $a_{\perp}$ and $a_{\parallel}$ denote the acceleration components that are perpendicular and parallel to velocity, respectively, with $a_{\parallel}>0$ for accelerating and $a_{\parallel}<0$ for decelerating. It then follows from  \eqref{eq:L3} that the load factor for banked level turn is related to the centrifugal acceleration as
$\kappa = \sqrt{1+a_{\perp}^2/g^2}$.
Together with \eqref{eq:Drag2} and \eqref{eq:F1}, we have
\begin{align}
&F=c_1V^2+\frac{c_2}{V^2}\left(1+\frac{a_{\perp}^2}{g^2}\right)+ma_{\parallel}.\label{eq:F2}
\end{align}

 Therefore, the instantaneous power required for the banked level turn flight with speed $V$, tangential acceleration $a_{\parallel}$, and centrifugal acceleration $a_{\perp}$ can be expressed as
\begin{align}
P_{\req}(V,a_{\parallel}, a_{\perp})&=|F| V
=\left | c_1 V^3+ \frac{c_2}{V}\left(1+\frac{a_{\perp}^2}{g^2}\right)+ma_{\parallel}V \right|.\label{eq:Preq2}
\end{align}
It is not difficult to verify that \eqref{eq:Preq} is a special case of \eqref{eq:Preq2} with $a_{\perp}=0$.

\subsubsection{Energy Required for Trajectory $\q(t)$}
For given velocity vector $\mathbf v$ and acceleration vector $\mathbf a$, the tangential and centrifugal accelerations can be obtained by decomposing $\mathbf a$ along the parallel and normal directions of $\mathbf v$, which gives
\begin{align}\label{eq:aDecomp}
&a_{\parallel}=\frac{\mathbf a^T \mathbf v}{\|\mathbf v\|}, \ a_{\perp}=\sqrt{\|\mathbf a\|^2-\frac{(\mathbf a^T \mathbf v)^2}{\|\mathbf v\|^2}}.
\end{align}
By substituting \eqref{eq:aDecomp} and $V=\|\mathbf v\|$ into \eqref{eq:Preq2}, the instantaneous power required  as a function of $\mathbf v$ and $\mathbf a$ can be expressed as
\begin{align}
P_{\req}(\mathbf v, \mathbf a)=\bigg| c_1\|\mathbf v\|^3+\frac{c_2}{\|\mathbf v\|}\bigg(1+\frac{\|\mathbf a\|^2-\frac{(\mathbf a^T \mathbf v)^2}{\|\mathbf v\|^2}}{g^2}\bigg)+m\mathbf a^T\mathbf v\bigg|.\label{eq:Preq3}
\end{align}

For an aircraft with trajectory $\q(t)$, and hence velocity $\mathbf v(t)=\dot{\q}(t)$ and acceleration $\mathbf a(t)=\ddot{\q}(t)$, the total propulsion energy required is then given by
\begin{align}
\bar E& (\q(t))=  \int_{0}^T P_{\req}\left( \mathbf v(t), \mathbf a(t)\right)dt,\label{eq:Eqt}
\end{align}
with $P_{\req}(\cdot)$ given by \eqref{eq:Preq3}.

For normal aircraft flight without drastic deceleration so that thrust reversal is not required, the term inside the  magnitude operator of \eqref{eq:Preq3} is always non-negative. As a result, \eqref{eq:Eqt} can be rewritten as
\begin{align}
 \hspace{-7ex} \bar E(\q(t))=&\int_{0}^T  \left[c_1\|\mathbf v(t)\|^3+\frac{c_2}{\|\mathbf v(t)\|}\bigg(1+\frac{\|\mathbf a(t)\|^2-\frac{(\mathbf a^T(t) \mathbf v(t))^2}{\|\mathbf v(t)\|^2}}{g^2}\bigg)\right]dt\notag \\
 &+ \int_0^T m\mathbf a^T(t)\mathbf v(t)dt.\label{eq:ER2}
\end{align}
Furthermore, the last term in \eqref{eq:ER2} can be expressed as
\begin{align}
\int_0^T m\mathbf a^T(t)\mathbf v(t)dt&=\int_0^T m\dot {\mathbf v}^T(t)\mathbf v(t)dt\\
&=\frac{1}{2}m\left[\|\mathbf v(T)\|^2-\|\mathbf v(0)\|^2\right],\label{eq:kinetic}
\end{align}
where the equality in \eqref{eq:kinetic} follows from the integral identity $\int \dot {\mathbf v}^T(t)\mathbf v(t)dt=\frac{1}{2}\|\mathbf v(t)\|^2$.


This thus completes the derivation of the UAV energy consumption model in \eqref{eq:ER}.

\section{Proof of Theorem~\ref{theo:energyMini}}\label{A:energyMini}
To obtain the optimal solution to problem \eqref{eq:powerMimum}, we first consider its relaxed problem by ignoring the equality constraints $\mathbf a(t)=\dot{\mathbf v}(t)$, $\forall t$, 
 whose optimal  value obviously gives a lower bound to that of problem \eqref{eq:powerMimum}. We then show that the optimal solution to this relaxed problem automatically satisfies the equality constraints of problem \eqref{eq:powerMimum}, and thus it is also optimal to \eqref{eq:powerMimum}.

 For the relaxed problem, it is evident  that the optimal solution should satisfy $\mathbf a^\star(t)=\mathbf 0$. Furthermore, by decomposing the velocity vector $\mathbf v(t)$ as $\mathbf v(t)=V(t)\vec{\mathbf v}(t)$, with $V(t)\geq 0$ and $\|\vec{\mathbf v}(t)\|=1$ denoting the time-varying speed and direction, respectively, the problem reduces to
 \begin{align}
\underset{V(t)}{\min}\ & \int_{0}^T  \left[c_1V^3(t)+\frac{c_2}{V(t)}\right]dt \notag \\
\text{s.t.}\ & V(t) \geq 0, \ \forall t. \label{eq:powerMimum2}
\end{align}

\begin{lemma}\label{eq:lemmaConstSpeed}
The optimum solution to problem \eqref{eq:powerMimum2} is
\begin{align}
V^\star(t)=V_{\emi}=\left(\frac{c_2}{3c_1} \right)^{1/4},\  \forall t.\label{eq:Vstar}
\end{align}
\end{lemma}

\begin{IEEEproof}
First, we show that constant speed is optimal  to problem \eqref{eq:powerMimum2}. To this end, define the function $f(V)=c_1V^3+c_2/V$, which can be shown to be a convex function for $V\geq 0$. For any time-varying speed $V(t)$ over a duration $T$, let $\bar V=\frac{1}{T}\int_0^T V(t)dt$ be the  average speed. By applying the Jensen's inequality to the convex function $f(V)$, we have
\begin{align}
f(\bar V) \leq \frac{1}{T} \int_{0}^T f(V(t))dt.
\end{align}
Thus, the average speed $\bar V$ always achieves a lower objective value to problem \eqref{eq:powerMimum2}. Thus, constant speed is optimal  to \eqref{eq:powerMimum2}, i.e., $V(t)=V, \forall t$. By setting the first-order derivative of $f(V)$ to zero, the optimal  speed in \eqref{eq:Vstar} can be obtained.
\end{IEEEproof}

As a result, the optimal  solution to the relaxed problem of \eqref{eq:powerMimum} is $\mathbf a^\star(t)=\mathbf 0$, $\mathbf v^\star(t)=(\frac{c_2}{3c_1})^{1/4}\vec {\mathbf v}(t)$, $\forall t$, with $\vec {\mathbf v}(t)$ being the arbitrary flying direction. Without loss of optimality, we may set $\vec {\mathbf v}(t)=\vec {\mathbf v}$, $\forall t$, i.e., by imposing a  time-invariant direction. As such, the equality constraint of problem \eqref{eq:powerMimum} is automatically satisfied since $\mathbf a^\star(t)=\dot{\mathbf v}^\star(t)=\mathbf 0$. This thus leads to the optimal solution in \eqref{eq:optEM}. The optimal  value in \eqref{eq:Emi} can then be obtained accordingly.

This completes the proof of Theorem~\ref{theo:energyMini}.

\section{Proof of Theorem~\ref{theo:LB}}\label{A:LB}
The proof of the lower bound in \eqref{eq:LB} follows similar arguments as that for Lemma~2 in \cite{641}. We first define the function $h(z)\triangleq \log_2\left(1+\frac{\gamma}{A+z}\right)$ for some constant $\gamma\geq 0$ and $A$, which can be shown to be  convex for $A+z\geq 0$. Using the property that the first-order Taylor expansion of a convex function is a global under-estimator \cite{202}, for any given $z_0$, we have $h(z)\geq h(z_0) + h'(z_0)(z-z_0)$, $\forall z$, where $h'(z_0)=\frac{-(\log_2 e) \gamma }{(A+z_0)(A+\gamma+z_0)}$ is the derivative of $h(z)$ at point $z_0$. By letting $z_0=0$, we have the following inequality,
  \begin{align}
 \log_2\left(1+ \frac{\gamma}{A+z}\right)
 \geq \log_2\left( 1+ \frac{\gamma}{A}\right) - \frac{(\log_2 e) \gamma z}{A(A+\gamma)}, \ \forall z. \label{eq:taylor}
 \end{align}
 Then for each time slot $n$, let $\gamma=\gamma_0$, $A=H^2+\|\q_j[n]\|^2$, $z=\|\q[n]\|^2-\|\q_j[n]\|^2$, the inequality in \eqref{eq:LB} thus follows.

Furthermore, it can be obtained from \eqref{eq:Rlb} that for any $n$, the gradient of $\bar R_{\lb}$ with respect to $\q[n]$ is
\begin{align}
\nabla_{\q[n]}\bar R_{\lb}=-2B\beta_j[n]\q[n], \ \forall n.\label{eq:grad1}
\end{align}
The gradient of $\bar R$ in \eqref{eq:LB} can be obtained as
\begin{align}
\nabla_{\q[n]}\bar R=\frac{-2B(\log_2 e)\gamma_0 \q[n]}{(H^2+\gamma_0+\|\q[n]\|^2)(H^2+\|\q[n]\|^2)}, \forall n.\label{eq:grad2}
\end{align}
The two gradients in \eqref{eq:grad1} and \eqref{eq:grad2} are identical when evaluated at $\q[n]=\q_j[n]$, $\forall n$.

This thus completes the proof of Theorem~\ref{theo:LB}.

\bibliographystyle{IEEEtran}
\bibliography{IEEEabrv,IEEEfull}

\end{document}